\documentclass{elsart}
\usepackage[lined,boxed,ruled]{algorithm2e}
\usepackage{times}

\usepackage[dvips]{graphicx}
\usepackage[usenames,dvipsnames]{color}
\usepackage{amssymb}
\usepackage{bez123}
\usepackage{amssymb}
\usepackage[square, comma, sort&compress]{natbib}

\usepackage[all]{xy}

\journal{Complexity}
\begin{document}
\begin{frontmatter}

\title{Simulating Network Influence Algorithms Using Particle-Swarms: PageRank and PageRank-Priors}
\author[LANL,UCSC,VUB]{Marko A. Rodriguez\corauthref{cor}}
\corauth[cor]{Corresponding author. (505) 606-1691}
\ead{marko@lanl.gov}
\author[LANL]{Johan Bollen}
\ead{jbollen@lanl.gov}
\address[LANL] { Digital Library Research \& Prototyping Team, Los Alamos National Laboratory \newline Los Alamos, New Mexico 87545}
\address[UCSC] {Computer Science Department, University of California at Santa Cruz \newline Santa Cruz, California 95064}
\address[VUB]  {Center for Evolution, Complexity, and Cognition, Vrije Universiteit Brussel \newline Brussel, Belgium 1050}

\begin{abstract}
A particle-swarm is a set of indivisible processing elements that traverse a network in order to perform a distributed function.  This paper will describe a particular implementation of a particle-swarm that can simulate the behavior of the popular PageRank algorithm in both its {\it global-rank} and {\it relative-rank} incarnations.  PageRank is compared against the particle-swarm method on artificially generated scale-free networks of 1,000 nodes constructed using a common gamma value, $\gamma = 2.5$.  The running time of the particle-swarm algorithm is $O(|P|+|P|t)$ where $|P|$ is the size of the particle population and $t$ is the number of particle propagation iterations.  The particle-swarm method is shown to be useful due to its ease of extension and running time.
\end{abstract}

\end{frontmatter}


\section{Introduction}

Influence, prestige, impact, and authority refer to a class of network metrics that utilize the structure of a graph, $G=\{N,E\}$, to derive an influence ranking, $\vec{I} \in \mathbb{R}^{|N|}$, over all its constituent nodes.  Generally these metrics determine a node's importance in a recursive fashion.  A node's influence, $\vec{I}_k$, is a function of the influence of the nodes that project to it.  This idea is represented in Eq. (\ref{eq:influence}), where $e_{j,k}$ is a directed edge from $n_j$ to $n_k$, $\mathrm{out}(n_j)$ is the set of outgoing edges from node $n_j$, and $t$ is the current iteration represented in discrete time.  The collection of influences across all nodes in the network is represented by the vector $\vec{I}$ which, upon convergence, is the principle eigenvector of the adjacency matrix formed by the graph \cite{google:brin1998}.\\

	\begin{equation}
		\label{eq:influence}
		\vec{I}_{k, t+1} = \sum_{\forall e_{j,k}\in E} \frac{{\vec{I}}_{j, t}}{|\mathrm{out}(n_j)|}
	\end{equation}
	
Since the inception of these algorithms there has been a strong focus on {\it global-rank}, $I(n|N)$ or simply $I(n)$, and only recently has there been research interest in {\it relative-rank} $I(n|R)$, where $R \subseteq N$ \cite{markov:white2003}. Global-rank determines the relative influence of each node with respect to the entire node population, $N$, while on the other hand, relative-rank determines the relative influence of each node with respect to a particular subset of the network, $R \subseteq N$.  Global-rank algorithms have found themselves at the forefront of web search techniques: PageRank \cite{google:brin1998}, HITS \cite{hits:kleinberg1999}, and their respective extensions.  Biased, or relative ranking has found application in domain-specific authority using web-page networks \cite{topic:haveliwala2002}, company-specific idea influence using collaboration networks \cite{markov:white2003}, and manuscript-specific peer-review influence using co-authorship networks \cite{peer:rodriguez2005}.  It is important to note that global-rank can be interpreted as a special case of relative-rank where each node's influence is calculated relative to a root node set that is the entire node population, $R = N$.\\

The contribution set forth by this paper is two fold.  First, this paper demonstrates the application of particle-swarms to the calculation of these two popular influence metrics: PageRank (global-rank) \cite{google:brin1998} and PageRank-Priors (relative-rank) \cite{markov:white2003}.  The particle-swarm algorithm is useful because of its running time and flexibility.  Unlike most popular implementations, a particle-swarm has a more tangible appeal that lends itself towards various functional modifications.  This paper will only provide the rudimentary data structures and functions necessary to simulate PageRank and PageRank-Priors, but the framework will provide room for possible extensions.  The second contribution of this paper is that it provides an introduction to the use of particle-swarms in the broader context of graph analysis and manipulation.  Currently there is little research in this area.  Of those manuscripts found, most of them analyze graphs from the perspective of a single random-walker and do no include more advanced functions and properties such as particle energy, decay, and teleportation \cite{brown:zhou2003, between:newman2003, random:tadic2003, random:noh2003}.\\

The outline of the paper is as follows.  {\it Section 2} will discuss both PageRank and PageRank-Priors from the standpoint of an object-oriented random-walker model.  {\it Section 3} will then describe the graph theoretic model of the particle-swarm method with emphasis on the various parameters and functions of the particles as they apply to simulating PageRank and PageRank-Priors. {\it Section 4} compares both PageRank algorithms and the particle-swarm algorithm on artificially generated scale-free networks.  Finally, {\it Section 5} discusses the running-time of the particle-swarm method and two optimizations.  The paper concludes, {\it Section 6}, with a short discussion of related PageRank algorithm implementations.

\section{Random-Walker Model}

Both PageRank \cite{google:brin1998} and PageRank-Priors \cite{markov:white2003} can be described in a random-walk fashion where a stochastic token, or particle, moves throughout a network, $G$.  The rank influence of any node $n_k \in N$ is the probability that the particle-token, $p$, will be seen at that node, $\vec{I}_k = P(p|n_k)$.  This conceptual analogy is explicitly represented within the object-oriented framework of this paper as a swarm of particle-tokens, $P$, that traverse the network landscape depositing their energy footprint on each node they traverse.  In doing so, the particles generate an influence ranking of the nodes in terms of the normalized energy distribution, $\vec{I}$, of the node population.

\subsection{PageRank Walker}
The PageRank algorithm, as described in \cite{google:brin1998}, was the driving force which has carried the Google search engine to the forefront of web search-engine technology.  Simply speaking, the algorithm is calculated in a recursive fashion where a particular page in a network of web-pages is influential if it is referenced by, or linked from, other influential pages.  Imagine a random-walker, $p$, traversing a network of web-pages such as the World Wide Web, $G=\{N,E\}$.  If that random-walker continuously finds itself at a particular page $n$, then that random-walker is said to have a high probability of being at that web page.  This probability is interpreted as the page's, or node's, influence.  The random-walker is consistently located at that web-page because the incoming edges to $n_k$, $\mathrm{in}(n_k) \subseteq E$, are either numerous, nearing the limit $|\mathrm{in}(n_k)| \approx |E|$, or the nodes that point to $n_k$ have a numerous set of incoming edges which allow the random-walker to consistently reappear at $n_k$.  Taken to its recursive limit, a node's influence is a measure of all the aggregate influence it receives from pages pointing to it whether direct or indirect. \\

A dampening-factor, $\lambda \in [0,1]$, can be introduced to reduce the spread of influence over time \cite{ranksinks:brin98}.  The further the random-walker travels, the less influence the random-walker should have, such that at full dampening, $\lambda = 1.0$, the random-walker can not take a step and all nodes are ranked equivalent, $\vec{I}_{k,t=0} = \frac{1}{|N|}$. The combination of random-walker propagation and dampening is expressed in Eq. (\ref{eq:influence2}).  The first block of the equation represents the proportion of influence distributed to $n_k$ by $n_j$.  This can also be interpreted as the probability of the random walker taking the edge $e_{j,k}$ given the condition that its current location is $n_j$. The second block of the equation provides the equal distribution of influence incurred through dampening.  Notice that $\lambda$ serves as the scaling variable modulating the influence of each block on the influence vector, $\vec{I}$.\\

	\begin{equation}
		\label{eq:influence2}
		\vec{I}_{k}=\left[\left(1-\lambda\right){\sum_{\forall e_{j,k}\in E}\frac{1}{|\mathrm{out}(n_j)|}}\right] + \left[\frac{\lambda}{|N|}\right]
	\end{equation}

\subsection{PageRank-Priors Walker}
The priors idea was first proposed by \cite{markov:white2003} in their formalization of a relative-rank extension to both PageRank and HITS.  Suppose the network data structure, $G=\{N,E\}$, is supplied with a root node set, $R \subseteq N$.  This root set is the set of nodes used to rank all other nodes relative too.  Suppose that at each time step, the random-walker has a probability, $\beta$, of 'teleporting' to particular node $r \in R$ as defined by the probability distribution, $\rho(r) = \frac{1}{\left|R\right|}$. This means that if the random walker decides to teleport home, which is dependent on the probability $\beta$, then the random walker chooses a random node in the root node set, $R$. A variation to the algorithm can bias the probability distribution over $R$.\\

As $\beta$ approaches 1.0 the probability of seeing the random-walker at any node in $R$ becomes greater and therefore the influence of the nodes in $R$, as well as those nodes that $R$ projects to, increases.  At the limit when $\beta = 1.0$, the influence distribution of all $n \notin R = 0.0$ and the influence of all $n \in R = \frac{1}{|R|}$.  In this way, the random-walker is biasing the ranking of the network nodes, $N$, towards the subset $R$.  When $\beta = 0.0$ there is still a bias towards the root node set since the random-walker will initiate its walk from that set, but the probability of the random-walker's location diffuses over the network as the amount of iterations increases.\\

The next section will extend the random-walker model to a particle-swarm model where a collection of random-walkers, $P$, traverse the network depositing an energy footprint at each step of the way.  These energy footprints, as stored in the node's 'memory', $\vec{I}_k \in \vec{I}$, represent the probability of having a particle at that particular node.  It is important to note that the random-walker model can be easily extended to account for weighted graphs, $G=\{N,E,W\}$, where the outgoing edges of a node are normalized to create a probability distribution.  This probability distribution biases the random-walkers decision when taking an outgoing edge and in such cases is called a biased random-walker.  In this way, weighted PageRank and weighted PageRank-Priors can be calculated.  The next section will discuss the full weighted model of the particle-swarm framework though the simulations are only for the PageRank and PageRank-Priors non-weighted counterparts.

\section{Particle-Swarm Model}

A particle-swarm, $P$, is a collection of unique processing entities that, by traversing a network in a stochastic manner, collectively perform a distributed function.  In relation to the random-walker model, a particle-swarm is simply a collection of many random-walkers.  The unification of the network particles, nodes, roots, edges, and weights form the data structure $G=\{P,N,R,E,W\}$ where each edge is assigned a weight, $|E|=|W|$, and $R \subseteq N$.  A single particle can contain any number of properties and behaviors, but for the purposes of this paper only those properties and behaviors that apply to PageRank and PageRank-Priors are described, $P = \{\epsilon, \delta, h, \beta, c\}$. A particle is an indivisible entity, but its local energy content, $\epsilon_i \in [0,1]$, is not.  Each time a particle traverses an edge, its local energy content is affected by a decay-scalar, $\delta_i \in [0,1]$, which is related to the dampening factor, $\lambda$, described previous.  To simulate PageRank-Priors a particle must have a reference to its originating, or root node, $h_i \in R$, so that it can 'teleport' home as determined by a back-probability, $\beta_i \in [0,1]$ and a back selection function $B(\beta_i) \in \{0, 1\}$. Finally, a particle traverses an outgoing edge from its current node location, $c_i \in N$, according to an edge selection function, $\theta(\mathrm{out}(c_i))$, which returns an edge $e_{i,j} \in \mathrm{out}(c_i)$.  These properties and functions are enumerated below for ease of reference. Note that in order to simulate PageRank and PageRank-Priors, $\delta$ and $\beta$ are the same for every particle in the simulations to follow, $\forall_{i,j} : \delta_i = \delta_l$ and $\beta_i = \beta_l$.  An obvious extension to this framework is to assign unique $\delta$ and $\beta$ values to different particles.\\
	
	\begin{enumerate}
		\item $\epsilon$: a local energy value $\epsilon \in [0,1]$
		\item $\delta$: a energy decay-scalar $\delta \in [0,1]$
		\item $h$: a reference to its home, or root, node $h \in R$
		\item $\beta$: a back-probability $\beta \in [0,1]$
		\item $c$: a reference to the current node location $c \in N$ 
		\item a probabilistic back selection function $B(\beta) \in \{0,1\}$
		\item a probabilistic outgoing edge selection function $\theta(\mathrm{out}(c))$ returns $e_{i,j} \in \mathrm{out}(c)$
	\end{enumerate}
	
A network node, $n_k$, is represented by the triplet $\{P(n_k), \mathrm{out}(n_k), \vec{I}_k\}$, where $P(n_k)$ is a unique set of particles located at $n_k$, $\mathrm{out}(n_k)$ is a unique set of outgoing edges from $n_k$, and $\vec{I}_k \in \mathbb{R}$ is $n_k$'s local energy value.  Any edge in the network, $e_{k,j}$, is a directed edge, from $n_k$ to $n_j$, with an associated weight, $w_{k,j} \in [0,1]$.  The weights of the set of all outgoing edges from any node, $\mathrm{out}(n_k)$, must be normalized to create a probability distribution for each particle's propagation function (Eq. \ref{eq:normedges}).\\

	\begin{equation}
		\label{eq:normedges}
			w_{{k,j}_{(t+1)}} = \frac{w_{{k,j}_{(t)}}}{\sum^{|\mathrm{out}(n_k)|}_{i=0}w_{{k,i}_{(t)}}}
	\end{equation}

Initially, a set of nodes in the network are seeded with a collection of particles, $P$. The initial particle distribution, $P$, can be an equal distribution or a biased distribution depending on the desired functional output.  For global-rank metrics, each node in the network is provided with an equal initial distribution, $|P(n_k)| = \frac{|P|}{|N|}$, while for relative-rank methods, only an initial root set, $R \subseteq N$, will be provided with particles, $|P(r_k)| = \frac{|P|}{|R|}$ where $r_k \in R$. \\

At each time step of the algorithm, a particle performs three behaviors.  First, the particle increments its current node's energy content, $\vec{I}_k$, with its current energy content, $\epsilon_i$, by way of $\vec{I}_{k,t+1} = \vec{I}_{k,t} + {\epsilon_i}_{(t)}$ (Alg. \ref{alg:page}-16).  Next, the particle decays its energy content by the parameterized decay-scalar, $\delta_i$ (Eq. \ref{eq:pathdecay}, Alg. \ref{alg:page}-17).\\
	
	\begin{equation}
		\label{eq:pathdecay}
 		{\epsilon_i}_{(t+1)} = {\epsilon_i}_{(t)} - (\delta_i{\epsilon_i}_{(t)})  
  \end{equation}

Lastly, the particle calculates $B(\beta)$ (Alg. \ref{alg:page}-18). If the function returns $1$, then the particle will return home, ${c_i}_{(t+1)} = h_i$. If the function returns $0$, then the particle chooses an outgoing edge of its local node, $e_{i,j} \in \theta(\mathrm{out}(c_i))$ (Alg. \ref{alg:page}-26).  The outgoing edge chosen, $e_{i,j}$, determines the particles new nodal reference, ${c_i}_{(t+1)} = n_j$.  A particle's death occurs when $\epsilon_i = 0.0$. Since the decay function of the particle is based on the percentage of its current energy content, formally the particle energy will approach, but never reach $0.0$.  Therefore, a threshold for particle death is given when $\epsilon_i \leq \vartheta$.  For the purposes of these simulations an arbitrarily low $\vartheta$ was chosen to be $10^{-8}$. Unlike the 'random teleport' functionality of most PageRank implementations, if node $c_i$ does not have an outgoing edge, then the particle is destroyed, $\epsilon_i = \vartheta$ (Alg. \ref{alg:page}-22). Once all the particles in the network have died or a desired $t$ has been reached the particle propagation algorithm is complete. The energy content, $\vec{I}_k$, of all nodes can be normalized to yield the proportion of energy every node has with respect to one another.  This proportion can be interpreted as the probability of seeing a random-walker at that particular node. The aggregated values of all energy in the network forms the influence vector $\vec{I}$.\\

The particle-swarm framework encapsulates both aspects of PageRank and PageRank-Priors while allowing for both implementations to be run in their original form.  For example, to simulate PageRank, $\beta = 0.0$, $\delta \in [0,1]$. To simulate PageRank-Priors, $\beta \in [0,1]$, $\delta = 0.0$.  A benefit of this framework is that hybrid algorithms can be implemented by combining back-probability, $\beta$, and energy decay, $\delta$, in the same simulation.\\

The pseudocode for the particle-swarm implementation of PageRank is provided in Alg. (\ref{alg:page}).  The first functional block expresses a particle-distribution algorithm and the second block expresses the particle-propagation algorithm. To implement PageRank-Priors the loop on line $3$ should run through $R$ not $N$ and a desired $\beta$ should be set at line $6$.  An overview of the different Big-O running times of the two functions are presented in their respective comments and will be examined more closely in the {\it Section 5}.\\

\incmargin{1cm}
\restylealgo{boxed}
\linesnumbered
\begin{algorithm}[h!]
\begin{footnotesize}
\Indp
		\CommentSty{\#distribute particles: $O(|N|particlesPerNode) = O(|P|)$}\;
		int $i = 0$\;
		\ForEach{$(n_k \in N)$}{
			int $particlesPerNode = 10$\;
			\For{$($$l=0$, $l<particlesPerNode$, $l$++$)$}{
				$\epsilon_i = 1.0$; $\delta_i = 0.15$; $h_i = n_k$; $\beta_i = 0.0$; $c_i = n_k$\;
				$i$++\;
			}
		}
		\BlankLine
		\CommentSty{\#disseminate particles: $O(|P|t)$}\;
		int $t = 0$\;
		\While{$(t < pageIterations)$}{
			$t$++\;
			\For{$($$i=0$, $i < |P|$, $i$++$)$}{
				\If{$($$\epsilon_i > \vartheta$$)$}
				{
					$\vec{I}_{c_i} = \vec{I}_{c_i} + \epsilon_i$\;
					$\epsilon_i = \epsilon_i - (\delta_i \ast \epsilon_i )$\;
					\If{$(B(\beta_i) == 1)$}{
						$c_i = h_i$\;}
					\Else{
						  \If{$($$|\theta(\mathrm{out}(c_i))| == 0$$)$}{
							$\epsilon_i = \vartheta$
						}
						\Else{
							$c_i = \theta(\mathrm{out}(c_i))$\;
						}		
					}
				}
			}
		}
	\label{alg:page}
	\caption{Particle-Swarm implementation of PageRank}
\end{footnotesize}
\end{algorithm}
\decmargin{1em}

The next section will provide simulation results of the aforementioned particle-swarm algorithm, with varying parameters.  The results of these simulations are compared to the results given by PageRank, PageRank-Priors, and In-Degree.

\section{Simulation Correlations}

This algorithm test suite was originally run on random networks and scale-free networks of a varying $\gamma \in [2.0, 3.0]$ and size $|N| \in [100,10000]$ with insignificant variation on the particle-swarm's simulation performance.  Since the network size and topology are not dimensions for analysis, only a collection of scale-free networks of $\gamma = 2.5$ and $|N| = 1000$ are used for the remainder of the paper.  For scale-free construction, each node is given a predetermined set size for their incoming connections as defined by Eq. (\ref{gammafunction}), where the random number $\psi \in [0,1]$, $|\mathrm{in}(n_k)| \leq |N|-1$, and $\mathrm{in}(n_k)$ is the set of incoming edges to $n_k$ \cite{gamma:aldana2003}.\\

	\begin{equation}
		\label{gammafunction}
			|\mathrm{in}(n_k)| = \lfloor \psi^{-[1.0 / (\gamma - 1.0)]} \rfloor
	\end{equation}

From here nodes randomly connect to one another until their maximum incoming connectivity is reached, at which point the network construction algorithm is complete.  By predetermining the maximum incoming connectivity of a node in this way, the topology of the network maintains a small portion of node hubs and a relatively large portion of sparsely connected nodes which is characteristic of many naturally occurring networks \cite{linked:barabasi2002}.

\subsection{In-Degree as a Trivial Case of PageRank and Particle-Swarm}

The trivial case of the random-walker model is when the random-walker is only allowed to take one step. This is a method for calculating the influence of a node with respects to In-Degree and is an extreme case of PageRank as $\lambda \rightarrow 1.0$ and $\delta \rightarrow 1.0$ or the algorithm is halted at $t=1$.  To simulate In-Degree, each edge in the network must be traversed at $t=1$. To accomplish this, every node is supplied with a collection of random-walkers proportional to its outgoing edge size, $|P(n_j)| = \alpha |\mathrm{out}(n_j)|$ where $\alpha \in \mathbb{N}^+$.  Now if each random-walker has an equal probability of taking any outgoing edge, then at $t=1$ the distribution of random-walkers across the set of nodes $N$ is the In-Degree influence of that node (Eq. \ref{eq:indegreeinfluence}).\\

	\begin{equation}
		\label{eq:indegreeinfluence}
		\vec{I}_k = \sum_{\forall e_{j,k}\in E} \frac{|P(n_j)|}{|out(n_j)|}
	\end{equation}

Since the set of all $e_{j,k} \equiv \mathrm{in}(n_k)$, then when substituting $|P(n_j)|$ for $\alpha |\mathrm{out}(n_j)|$ Eq. (\ref{eq:indegreeinfluence}) can be represented as $\vec{I}_k = \alpha |\mathrm{in}(n_k)|$.  This equation produces an influence calculation perfectly correlated to In-Degree.  Given that this is a probabilistic particle, stochastic noise will disrupt the probability that each outgoing edge of every node is taken once and only once.  As the size of the initial distribution of particles increases, as $\alpha$ increases, the noise is reduced and the appropriate In-Degree influence vector is returned.  If the distribution of random-walkers is equal, $|P(n_k)| = \frac{|P|}{|N|}$, then only an approximation of In-Degree can occur.  In such cases, the more uniform the distribution of outgoing edges of all the nodes, the more accurate the approximation.\\

\begin{figure}[h!]
	\centering
		\scalebox{0.5}[0.5]{
		\rotatebox{-90}{\includegraphics[width=0.71\textwidth]{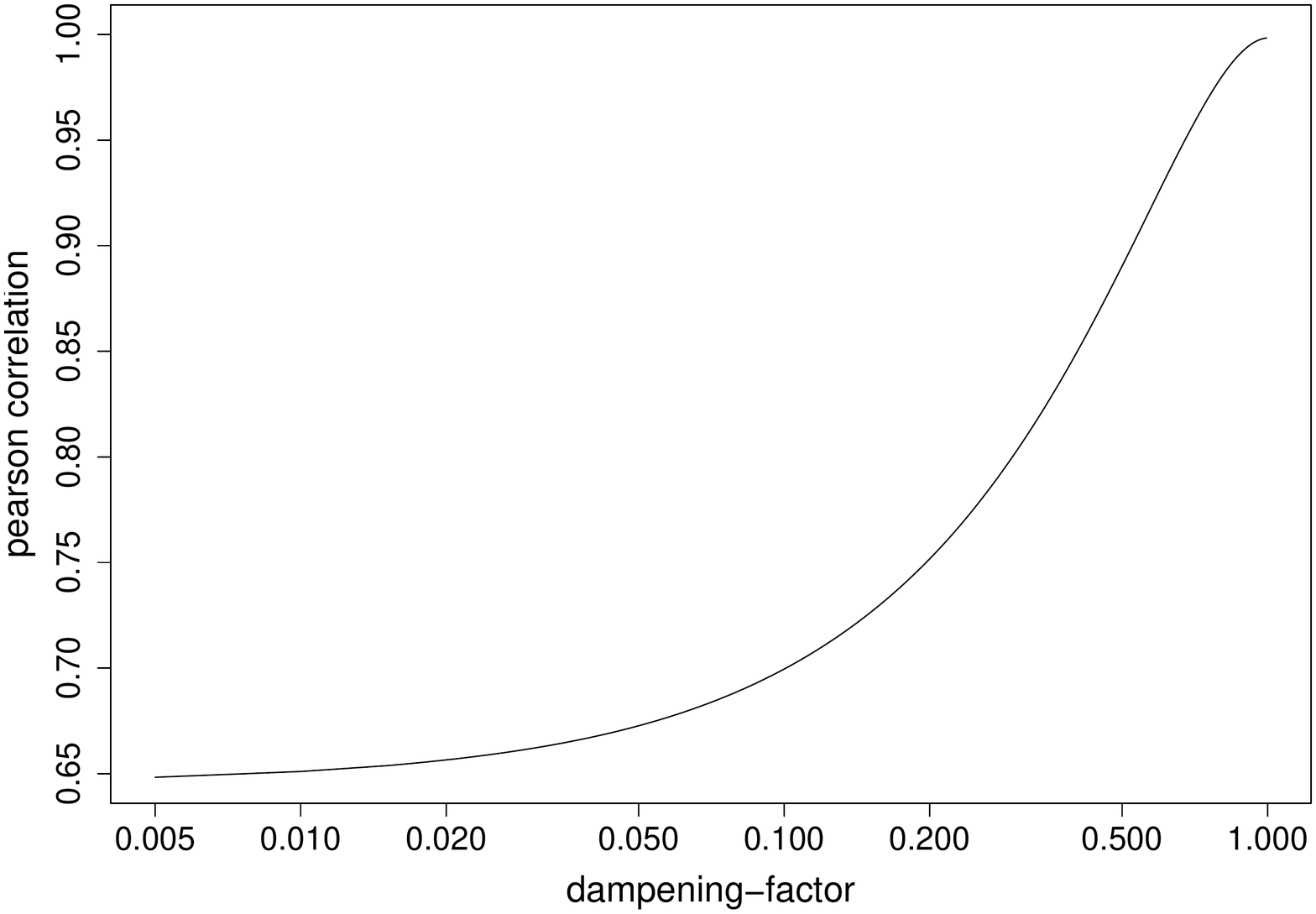}}
		\rotatebox{-90}{\includegraphics[width=0.71\textwidth]{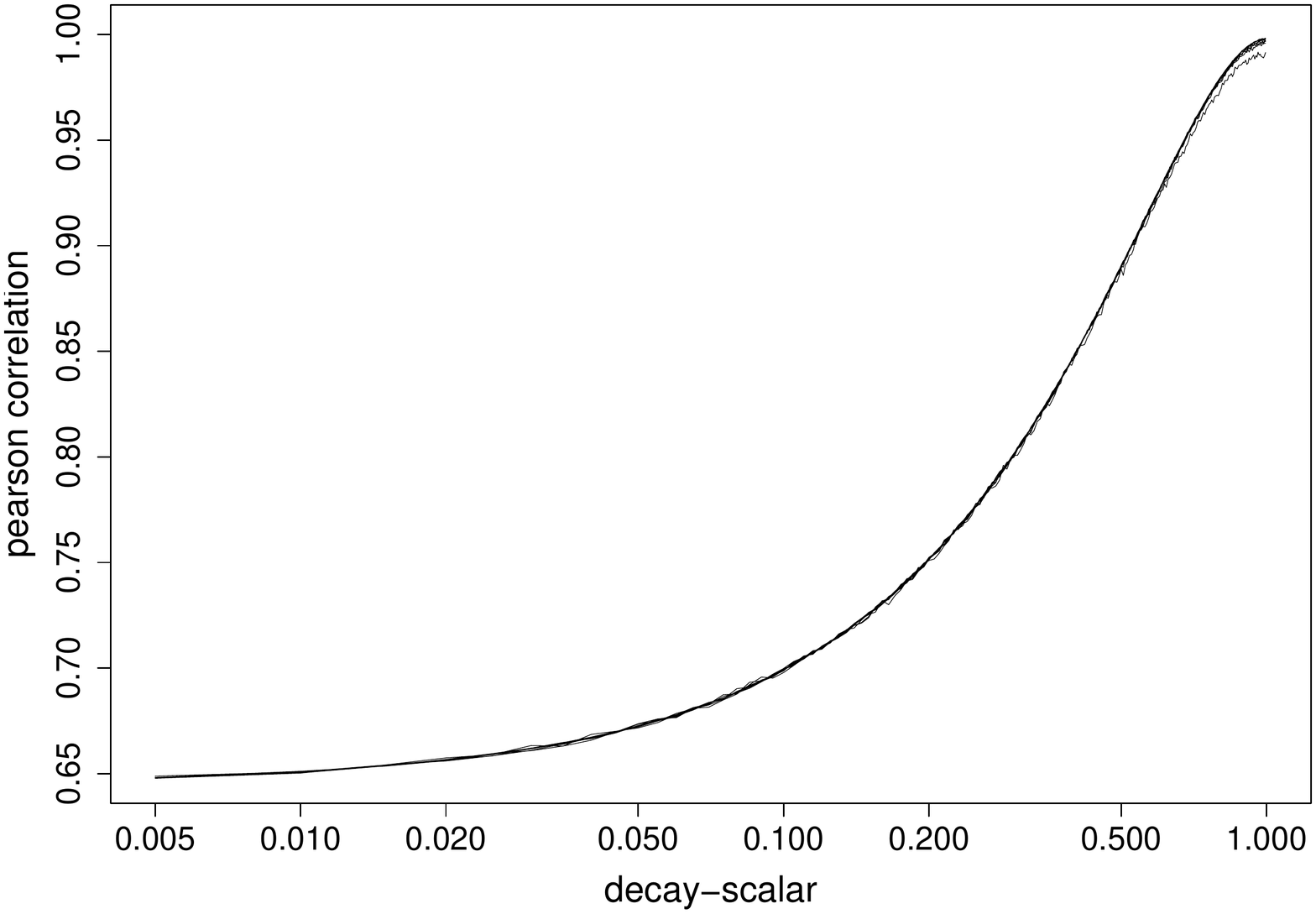}}}
	\caption{\textbf{a.} PR vs. IN over $\lambda \in [0.005, 0.995]$ \textbf{b.} PS vs. IN over $\delta \in [0.005, 0.995]$}
	\label{fig:pagein}
\end{figure}

Now that In-Degree has been described as a trivial case of PageRank and the particle-swarm method, both metrics will now approximate the In-Degree influence vector, $\vec{I}_{IN}$.  To simulate In-Degree influence using PagePank, $\lambda$ was scaled between $0.005$ and $0.995$ to produce the correlation plot, (Fig. \ref{fig:pagein}a). The reason for limiting the experiment to $\lambda = 0.995$ is because when $\lambda = 1.0$ there is no deviation in the rank vector, $\vec{I}_k=\frac{1}{|N|}$.  It is shown that PageRank best approximates In-Degree at the limit as $\lambda \rightarrow 1.0$, $C = 0.998$.  Next, the particle-swarm method for simulating In-Degree was determined using various initial particle distribution sizes of $|P(n_k)| \in [1,20]$, $|P| \in [1000, 20000]$, and $\beta = 0.0$. The $\delta$ of each particle was scaled from $0.005$ to $0.995$ and as $\delta \rightarrow 1.0$, In-Degree influence is approximated most closely, $C = 0.997$ (Fig. \ref{fig:pagein}b). Figure \ref{fig:pagein}b is composed of $20$ superimposed particle distribution size plots.  Note that the divergent plot in Figure \ref{fig:pagein}b occurs when $|P(n_k)|=1$, $|P|=1000$. The following influence vector relationship exists between these three algorithms: $\vec{I}_{\mathrm{IN}} \approx \vec{I}_{\lambda \rightarrow 1.0} \approx \vec{I}_{\delta \rightarrow 1.0}$.  Notice that PageRank and the particle-swarm method are nearly equivalent in their behavior for their respective $\delta = \lambda$ values, $\vec{I}_{\lambda} \approx \vec{I}_{\delta}$ when $|P(n_k)| \geq 1$.\\

\subsection{Correlating Particle-Swarm to PageRank and PageRank-Priors}

\begin{figure}[h!]
	\centering
		\scalebox{0.5}[0.5]{
		\rotatebox{-90}{\includegraphics[width=0.73\textwidth]{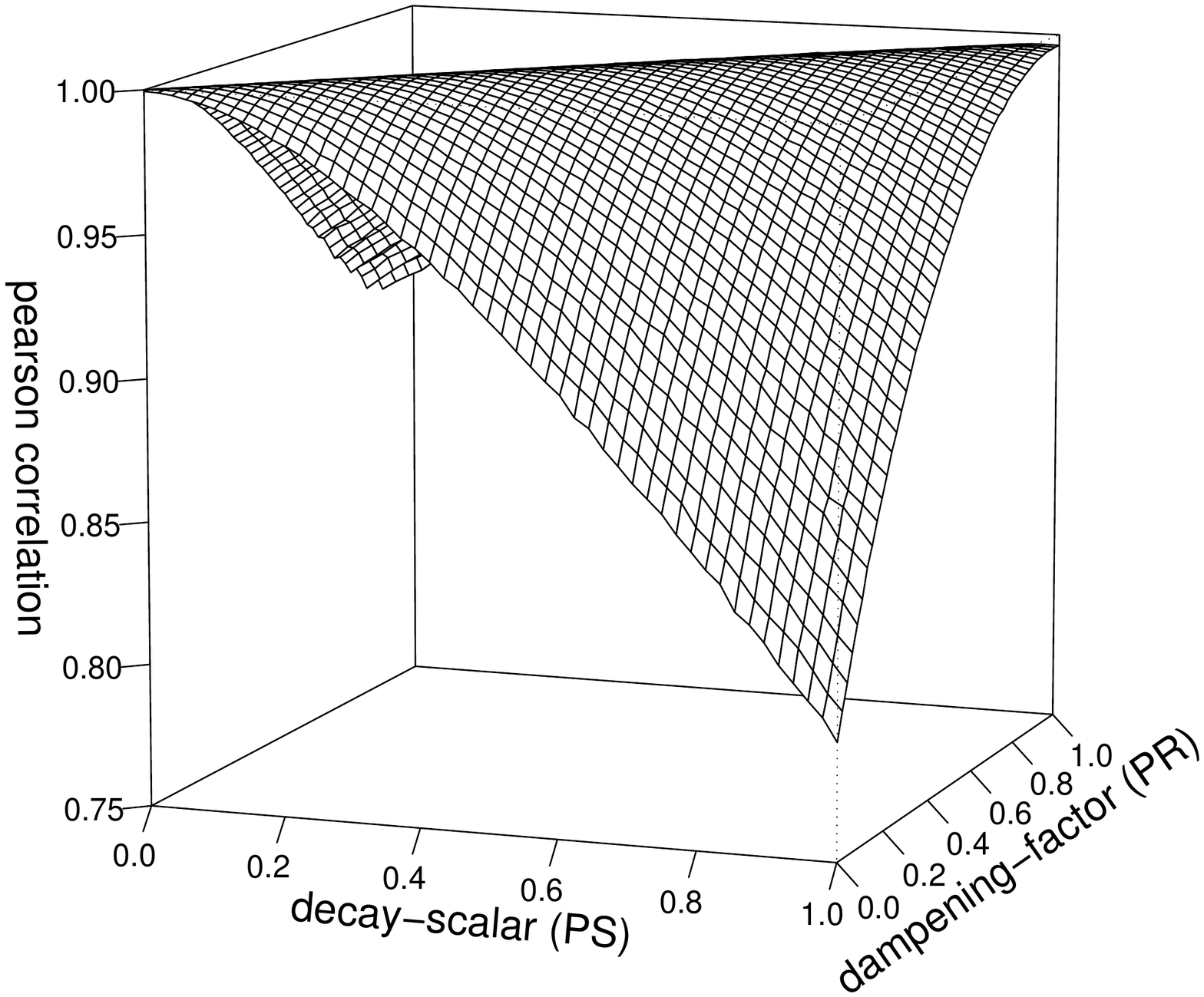}}
		\rotatebox{-90}{\includegraphics[width=0.73\textwidth]{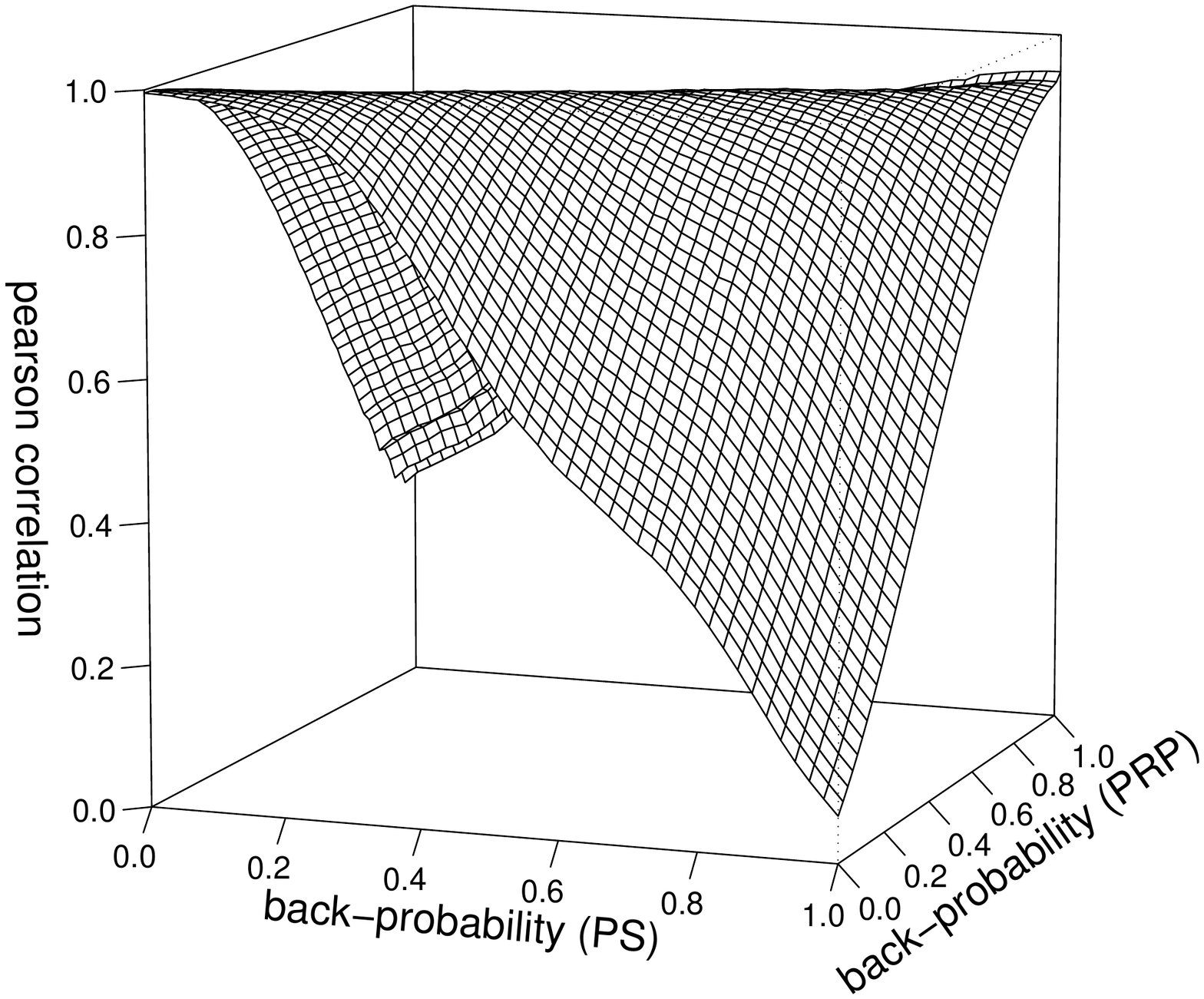}}}
	\caption{\textbf{a.} PR vs. PS over $\delta$ and $\lambda$  \textbf{b.} PRP vs. PS over $\beta$}
	\label{fig:pagerank}
\end{figure}

To simulate the results of PageRank (global-rank), the decay-scalar $\delta$ was varied between $0.005$ and $0.995$ for every potential dampening factor $\lambda$ between $0.005$ and $0.995$. The iterations of the particle-swarm method were constrained to $t_{PS}=t_{PR}$, where $t_{PS}$ and $t_{PR}$ are the amount of iterations for the particle-swarm method and PageRank, respectively.  Note that when $\delta$ is high, particle death can occur before the amount of iterations is complete.  For this experiment $|P(n_k)|=10$, $|P|=10000$. Figure \ref{fig:pagerank}a shows that an equal distribution of particles across all of $N$ with $\beta = 0.0$ simulates the respective PageRank calculation with a near $1.0$ Pearson correlation when $\delta = \lambda$.\\

PageRank-Priors (relative-rank), on the other hand, is a function of two variables, the size of the root node set, $R$, and the back-probability, $\beta$. The root node set was determined by randomly assigning a portion of the node population to $R$, $R = f(N,\varphi)$ where the percentage $\varphi \in [0.01,1.0]$ and $|R|=\varphi|N|$.  The selection of the root node set had limited effect on the correlation between PageRank-Priors and the particle-swarm method. Therefore, to represent the correlations in a 3D plot, the $\varphi$ factor was omitted (Fig. \ref{fig:pagerank}b).  The iterations of the particle-swarm method were constrained to $t_{PS}=t_{PRP}$ where $t_{PRP}$ is the amount of iterations required for PageRank-Priors to converge.  Furthermore, $\delta = 0.0$ since PageRank-Priors has no dampening-factor parameter. Figure \ref{fig:pagerank}b provides the correlation values of the particle-swarm's $\beta_{\mathrm{PS}} \in [0.1,1.0]$ for all $\beta_{\mathrm{PRP}} \in [0.1,1.0]$ of PageRank-Priors.  The root node set was generated from $10$\% of the node population, $\varphi = 0.10$, therefore when $|P(r_k)| = 10$, $|P| = 1000$.  PageRank-Priors and the particle-swarm method are strongly correlation when $R_{\mathrm{PRP}} = R_{\mathrm{PS}}$ and $\beta_{\mathrm{PRP}} = \beta_{\mathrm{PS}}$.\\

Any variation in the influence vectors between the particle-swarm method and PageRank-Priors is due in part to particle death when $|out(c_i)| = 0$ (Alg. \ref{alg:page}-22).  Since PageRank-Priors models a random walker's home return as a jump to any node in $R$, then all nodes in $R$ have an equal probability of being jumped to (assuming $\rho(r) = \frac{1}{|R|}$). On the other hand, a particle, when returning home, returns to its initial destination, $c_i = h_i$ (Alg. \ref{alg:page}-19).  If a particular outgoing path from an initial node is atrophied, then the potential for $|out(c_i)| = 0$ is greater and the potential for $\rho(r) = \frac{1}{|R|}$ is less.  Even at $\beta=1.0$, particle death is still a possibility. The rationale for designing the particle framework in this manner is to ensure decentralization for extended applications of the particle-swarm method. No particle has knowledge of $R$, only its particular initial, or root, node, $h_i$.\\

\section{Optimizations and Running Time}

This section will extend the current particle-swarm model to express two optimizations: {\it iteration constraining} and {\it random seeding}.  Currently, the running time of the particle-swarm method is $O(|P|+|P|t)$ where $|P|$ is the number of particles used in the simulation, and $t$ is the number of particle propagation iterations. In comparison, the running time of both PageRank and PageRank-Priors is $O(|E|t)$ where $E$ is the set of edges in the network and $t$ is the number of iterations required for convergence \cite{inside:bianchini2005, jung}.  It is important to note that $|P|$ is a function of $|N|$, not $|E|$, and for most real-world networks $|N| << |E|$.  An accurate particle-swarm simulation of PageRank is possible when $|P(n_k)| = 1$ and therefore $|P| = 1000$. For a $\gamma = 2.5$ scale-free network of $1000$ nodes $|E| \approx 2575$.  Therefore, the Big-O speed up, given $20$ iterations for each algorithm, is a factor of approximately $2.45=\frac{(2575)(20)}{(1000) + (1000)(20)}$, $\frac{|E|t}{|P| + |P|t}$.\\

Greater gains are seen in the particle-swarms simulation of PageRank-Priors when $|R| < |N|$. The particle population of a node is a proportion of the total population, $|P(r_k)| = \frac{|P|}{|R|}$. This ratio allows for a smaller particle population to be used when simulating PageRank-Priors without degrading the accuracy of the calculation. Notice that $|P(r_k)| = \frac{|P_{\mathrm{PRP}}|}{|R|} = \frac{|P_{\mathrm{PR}}|}{|N|}$, where $P_{\mathrm{PRP}}$ and $P_{\mathrm{PR}}$ are the particle sets for PageRank-Priors and PageRank, respectively. For $|P| = |R|$, the particle-swarm algorithms has a running time of $O(|R|+|R|t)$ when simulating PageRank-Priors. The PageRank-Priors particle-swarm simulation is more efficient in terms of running time than its originally, and only, published analysis of $O(|E|t)$ \cite{markov:white2003}. The benefits of the particle-swarm simulation of PageRank-Priors are best realized when $|R|<<|N|<<|E|$.\\

These calculations assume that the particle-swarm method and PageRank/PageRank-Priors both share the same amount of iterations, $t_{PS} = t_{PR}$, and that the particle-swarm method has a homogeneous initial particle seeding of at least $1$ particle per node. Both of these parameters can be reduced to lower the particle-swarms running time with varying effects on the correlation.  The following list of variables will be discussed in the following subsections and are presented here for ease of reference.\\
	\begin{enumerate}
		\item $t_{PS}$: number of iterations to propagate particles $t_{PS} \in \mathbb{N}^+$
		\item $\phi$: proportion of nodes to receive an initial seeding of particles $\phi \in [0,1]$
		\item $\alpha$: number of particles per node in the initial seeding $\alpha \in \mathbb{N}^+$
		\item $S$: the set of nodes receiving particles from the initial seeding $S \subseteq N$ and $|S|=\phi|N|$
	\end{enumerate}
	
\subsection{Constraining Particle Iterations and Random Particle Seeding}

\begin{figure}[h!]
	\centering
		\scalebox{0.5}[0.5]{
		\rotatebox{-90}{\includegraphics[width=0.71\textwidth]{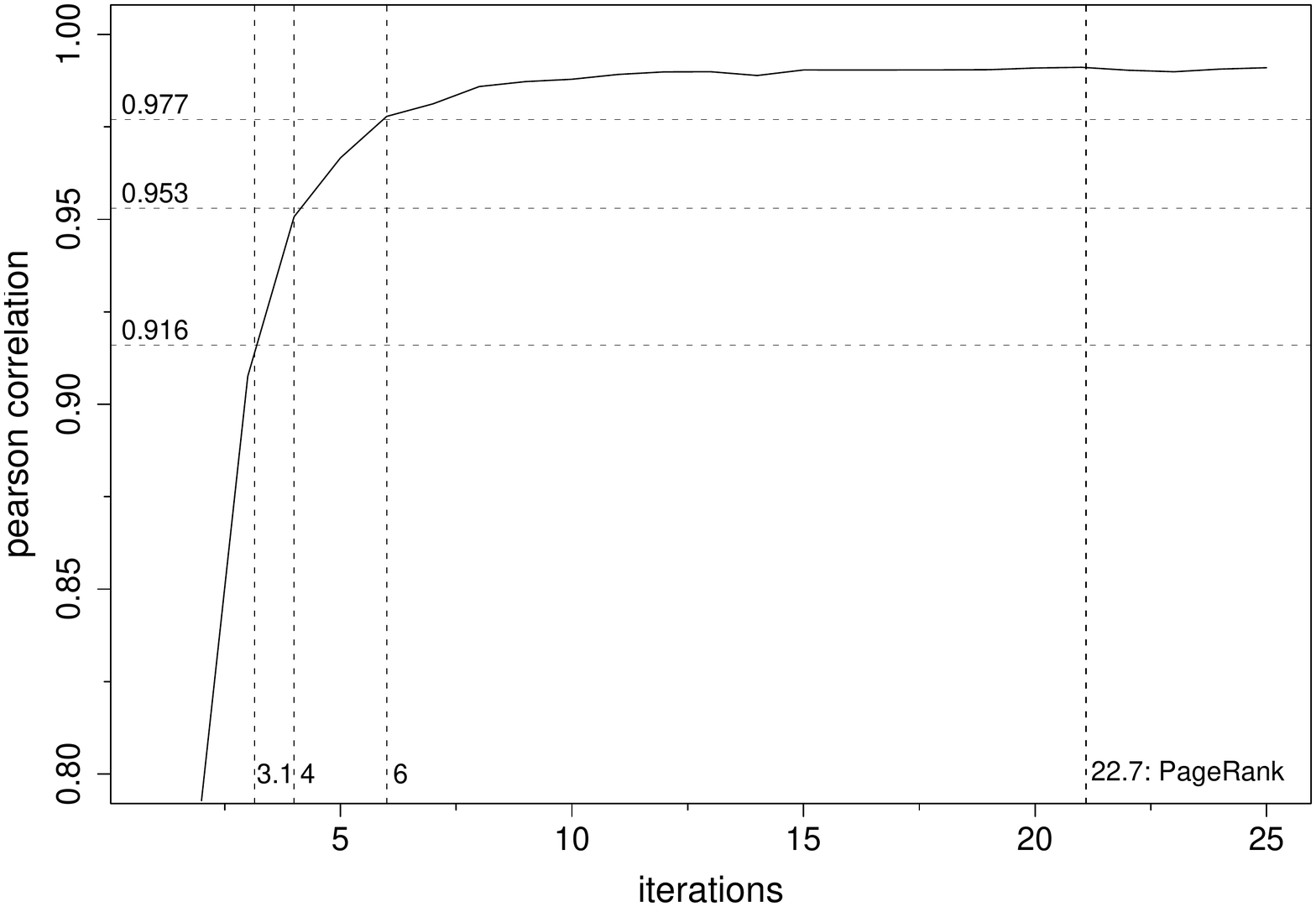}}
		\rotatebox{-90}{\includegraphics[width=0.71\textwidth]{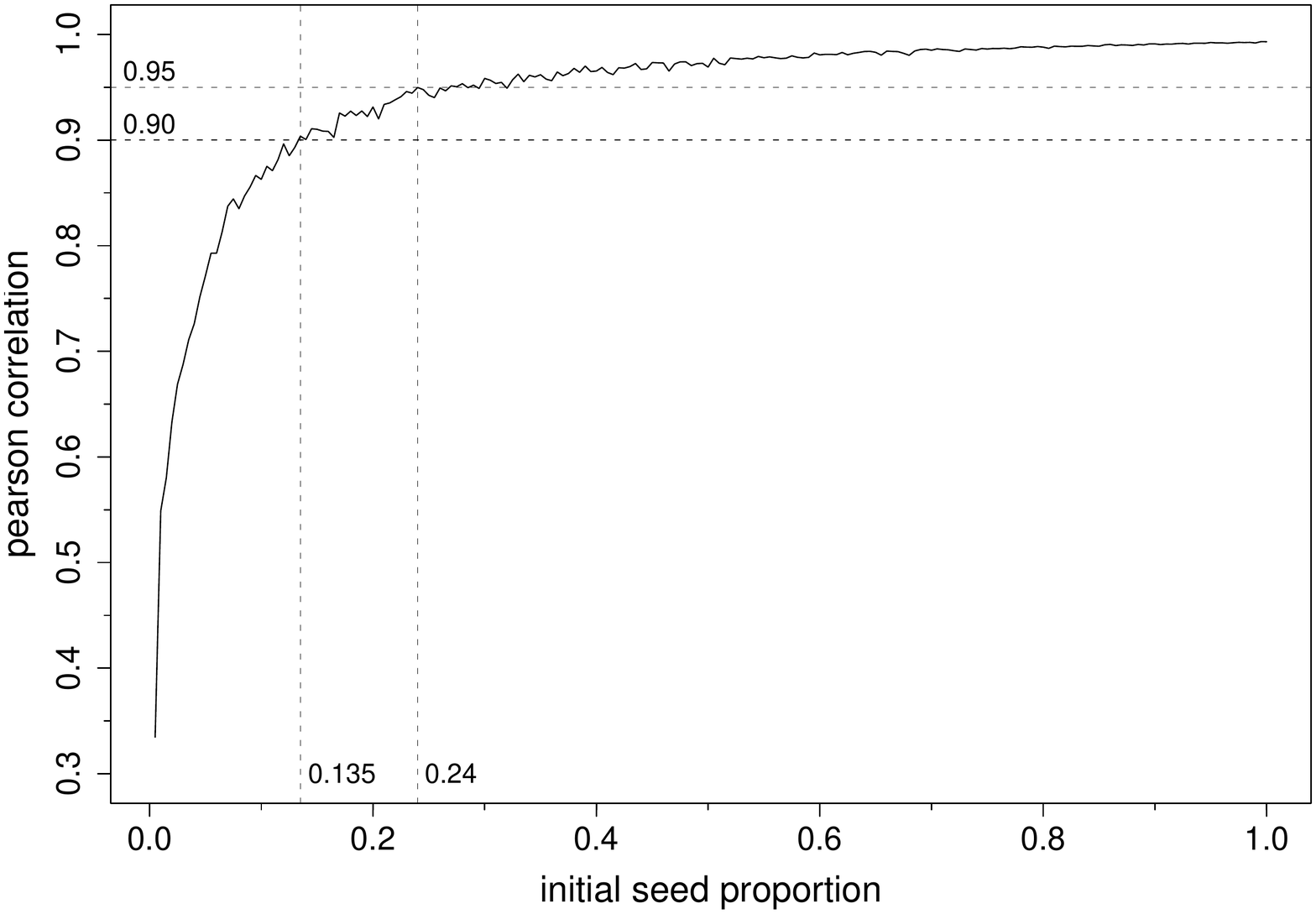}}}
	\caption{\textbf{a.} PS vs. PR over $t \in [1,25]$ \textbf{b.} PS vs. PR over $\phi \in [0,1]$}
	\label{fig:optimize}
\end{figure}

Algorithm \ref{alg:page}-12 assumes that a particle propagates for the same amount of iterations as PageRank, $t_{PS} = t_{PR}$.  This is a costly method since, to determine $t_{PR}$, PageRank must be executed. Another way of determining the amount of iterations for the particle-swarm method is to wait until all particles have died, which occurs when the particle's energy content has decayed to $\epsilon_i = \vartheta$ or when $c_i$ no longer has outgoing edges.  For a $\delta = 0.15$ and when $c_i$ always has at least one outgoing edge, particle death occurs after $113$ iterations ($\vartheta = 10^{-8})$, while the average PageRank converges after $22.7$ iterations on a $\gamma = 2.5$ scale-free network.  This obviously is not the fastest method either.  Therefore, Figure \ref{fig:optimize}a plots the correlation between the particle-swarm method and PageRank as the particle-swarm method's iteration value is constrained, $t_{PS} \in [1,25]$.  The range from $25 < t_{PS} \leq 113$ is omitted due to insignificant variation in the algorithm's behavior.  The result demonstrates that the particle-swarm method is strongly correlated with PageRank, $C=0.953$, after only $4$ iterations, $t_{PS}=4$.\\

Given different gamma values, the amount of iterations should vary.  For example, a $\gamma = 2.0$ scale-free network with $|N| = 1000$ only requires $12.52$ iterations for PageRank to converge.  Similarly, The particle-swarm method requires only $3.01$ iterations to produce a $C \approx 0.95$.  At the other extreme, a $\gamma = 3.0$ scale-free network requires approximately $28.88$ iterations to converge while the particle-swarm method requires $6.23$ iterations. The general trend, though non-linear, for producing a $C \approx 0.95$, is $t_{PS} \approx \frac{1}{4}t_{PR}$ or for each $\gamma \in [2.0, 3.0]$, $t_{PS} \approx 2\gamma$.\\

The particle-swarm method can also be optimized by randomly choosing a subset of the network to initially seed with particles, $S \subseteq N$. This random subset can be expressed as a proportion of the whole network, $\phi|N|$ where $\phi \in [0,1]$ and $|S|=\phi|N|$. Figure \ref{fig:optimize}b plots the correlation between PageRank and the particle-swarm method for different initial particle seed proportions.  It is shown that at $\phi=0.24$, when only $24\%$ of the nodes in the network are seeded with a single particle, the Pearson correlation is approximately $0.95$.  Therefore, an accurate PageRank calculation does not require all nodes to begin with an equal set of particles.  Thus, $|P| << |N|$.

\subsection{Combining the Optimizations}

\begin{figure}[h!]
	\centering
		\rotatebox{-90}{\includegraphics[width=0.8\textwidth]{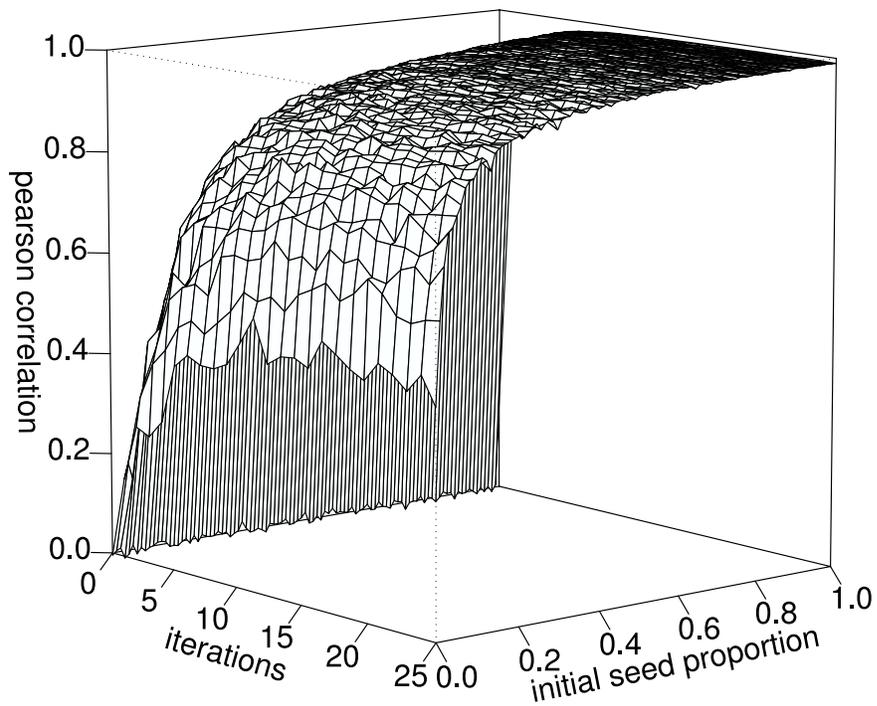}}
	\caption{PS vs. PR where PS combines iteration constraining and random seeding}
	\label{fig:combine}
\end{figure}

\begin{figure}[h!]
	\centering
		\rotatebox{-90}{\includegraphics[width=0.65\textwidth]{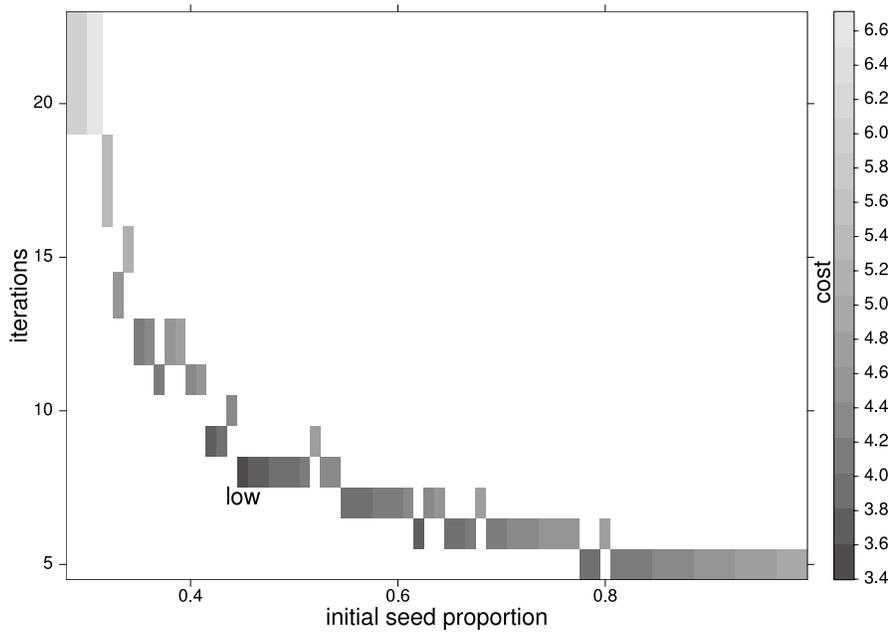}}
	\caption{$\phi t$ (cost) for various iteration/seed proportion pairs at $C \approx 0.95$}
	\label{fig:above95}
\end{figure}

The combination of both optimizations is represented in Figure \ref{fig:combine} where each initial seed proportion, $\phi \in [0.01,0.5]$, is calculated for every iteration amount, $t_{PS} \in [1,25]$.  Next, Figure \ref{fig:above95} plots the iteration amount against the seeding proportion for the lowest value pair obtaining a $C \approx 0.95$.  Each plot point's shade value is calculated as $\phi t_{PS}$, which represents the cost of performing that parameter pair.  Therefore, to achieve a $C \approx 0.95$, the most computationally efficient way is to use single particles ($\alpha = 1$), initially distributed over a moderate amount of nodes ($\phi \approx 0.45$), and propagated over a moderate amount of time steps ($t_{PS} \approx 8$).\\

The speed-up of the particle-swarm method with respects to PageRank is represented in Eq. (\ref{eq:speedup}) as $\Phi$. Since $\phi|N|\alpha$ represents the particle population, the full running time can still be expressed as $O(|P|+|P|t_{PS})$. The numerator in Eq. (\ref{eq:speedup}) is based on the standard PageRank implementation of $O(|E|t_{PR})$ \cite{inside:bianchini2005, jung}.\\

	\begin{equation}
		\label{eq:speedup}
		\Phi = \frac{|E|t_{PR}}{\left(\phi|N|\alpha\right) + \left(\phi|N|\alpha\right)t_{PS}}
	\end{equation}

For a $\gamma = 2.5$ scale-free network of $|N|=1000$, the theoretical speed-up of the fastest particle-swarm method yielding a $C \approx 0.95$ ($\alpha=1$, $\phi=0.45$, $t=8$) is calculated to be $14.43 = \frac{(2575)(22.7)}{\left[(0.45)(1000)(1)\right]+\left[(0.45)(1000)(1)(8)\right]}$.  To verify this hypothesis, PageRank, as implemented in \cite{jung}, was compared against the most optimized particle-swarm method. The benchmark testing was done over $500$ trials of $500$ different $\gamma=2.5$ scale-free networks of $|N|=1000$ with the average speed-up factor determined to be $22.23$.  A potential explanation for the increased benchmark speed-up relative to the theoretical speed-up may be in part to the fact that over the course of the particle-swarm algorithm, particles die before all iterations are complete (Alg. \ref{alg:page}-15,17,23).  Therefore, the general rule is that as $t$ increases, $|P|$ decreases.\\

\section{Conclusion}

Due to the popularity of the the global-rank implementation of PageRank there exists much literature on efficient implementations of the algorithm. One particular example includes an algorithm that partitions the graph into related influence clusters \cite{pagegraph:broder2004}. The graph clustering method groups nodes of a similar PageRank into a hyper-node and then calculates the full converging PageRank vector on the newly constructed hyper-network.  In this way, the clustering method is able to reduce the total amount of edges, $E$, iterated over.  The publication states that the typical edge reduction between the original network and the hyper-network is a factor of $20$ for networks containing billions of edges. The paper states a Spearman correlation of $0.95$ and a $2$ fold increase in calculation time relative to a 'highly optimized' implementation of PageRank. Edge reduction, by way of node grouping, also reduces the amount of nodes in the networks.  Therefore, there is a strong incentive to combine the graph clustering method and the particle-swarm method. This has not been tested as of yet.\\

Finally, the space constraints of the particle-swarm method are larger than traditional matrix methods since these methods do not represent particles, only the influence vector, $\vec{I}$, and the adjacency matrix of the graph.  This representation lends itself towards efficient space modifications \cite{spacepage:haveliwala2000}. The particle-swarm implementation discussed in this paper is calculated solely in main memory for small networks less than 10,000 nodes.  This test-bed implementation is obviously not useful for calculations on web-sized networks.  Future work will describe a system architecture for performing particle-swarm algorithms on large-scale networks.\\

The particle-swarm method for graph analysis has an appeal in its potential for functional modification. From the object-oriented perspective, a particle can be seen as an 'agent' that can contain any number of properties and behaviors.  The potential for modifying the particle-swarm framework presented in this paper can lead to a host of augmentations to the mentioned influence metrics.  One example includes the incorporation of 'negative' energy particles to reduce specific node influence as explained in \cite{peer:rodriguez2005}. New particle-swarm metrics are currently being implemented and results will be presented in future publications. This paper's simulations were performed using the Confluence package \cite{conf:rodriguez}. Our Confluence API has been written such that new particles can be easily extended to the basic 'energy' particle framework.

\section{Acknowledgments}

The use of particle-swarms for network analysis was first introduced to the first author by Daniel Steinbock and then later applied in a joint paper on distributing voting influence within trust-based social-networks \cite{ddd:rodriguez2004}.  Thanks to Carlos Gershenson and Francis Heylighen for many discussions on this topic.  Finally, a special thanks to the producers of JGraph (www.jgraph.com) and Scott White's Java implementation of PageRank and PageRank-Priors.  This work was partially funded by a GAANN Fellowship from the U.S. Department of Education and supported by Los Alamos National Laboratory.

 		\bibliographystyle{elsart-num}
		\bibliography{markobib}

\end{document}